\documentclass[a4paper]{jpconf}
\usepackage{amsmath}
\usepackage{graphicx}
\usepackage{wrapfig}
\usepackage{epsfig}

\begin{document}

\title{UNEDF: Advanced Scientific Computing Transforms the Low-Energy Nuclear Many-Body Problem}

\author{M. Stoitsov$^{1,2}$,  H. Nam$^{2}$,  W. Nazarewicz$^{1,2}$,
A. Bulgac$^{3}$, G. Hagen$^{2}$, M. Kortelainen$^{1,2}$, J. C. Pei$^{1,2}$, K. J. Roche$^{4}$, N. Schunck$^{5}$, I.~Thompson$^{5}$, 
J.P. Vary$^{6}$,  S. M. Wild$^{7}$
}

\address{
$^1$ Department of Physics and Astronomy, University of Tennessee Knoxville, TN 37996, USA \\
$^2$ Oak Ridge National Laboratory, P.O. Box 2008, Oak Ridge, TN 37831, USA \\
$^3$ Department of Physics, University of Washington, Seattle, WA 98195-1560, USA  \\
$^4$ Pacific Northwest National Laboratory, Richland, WA 99352, USA \\
$^5$ Lawrence Livermore National Laboratory, L-414, P.O. Box 808, Livermore, CA   94551, USA \\
$^6$ Department of Physics and Astronomy, Iowa State University, Ames, IA 50011-3160, USA \\
$^7$ Math \& Computer Science Division, Argonne National Laboratory, Argonne, IL 60439, USA 
}

\ead{stoitsovmv@ornl.gov}

\begin{abstract}
The UNEDF SciDAC collaboration of nuclear theorists, applied mathematicians, and computer scientists is developing a comprehensive  description of nuclei and their reactions that delivers maximum predictive power with quantified uncertainties.  This paper illustrates significant milestones accomplished by UNEDF through integration  of the theoretical approaches,  advanced numerical algorithms, and leadership class computational resources.
\end{abstract}

\vspace{-30pt}
\section{Introduction}

The frontier of discovery in nuclear physics aims to explain the nature of nuclei which is an essential component in energy, medical, and biological research, and national security.
\begin{wrapfigure}{l}{0.35\textwidth}
\vspace{-20pt}
\begin{center}
\includegraphics[width=0.35\textwidth]{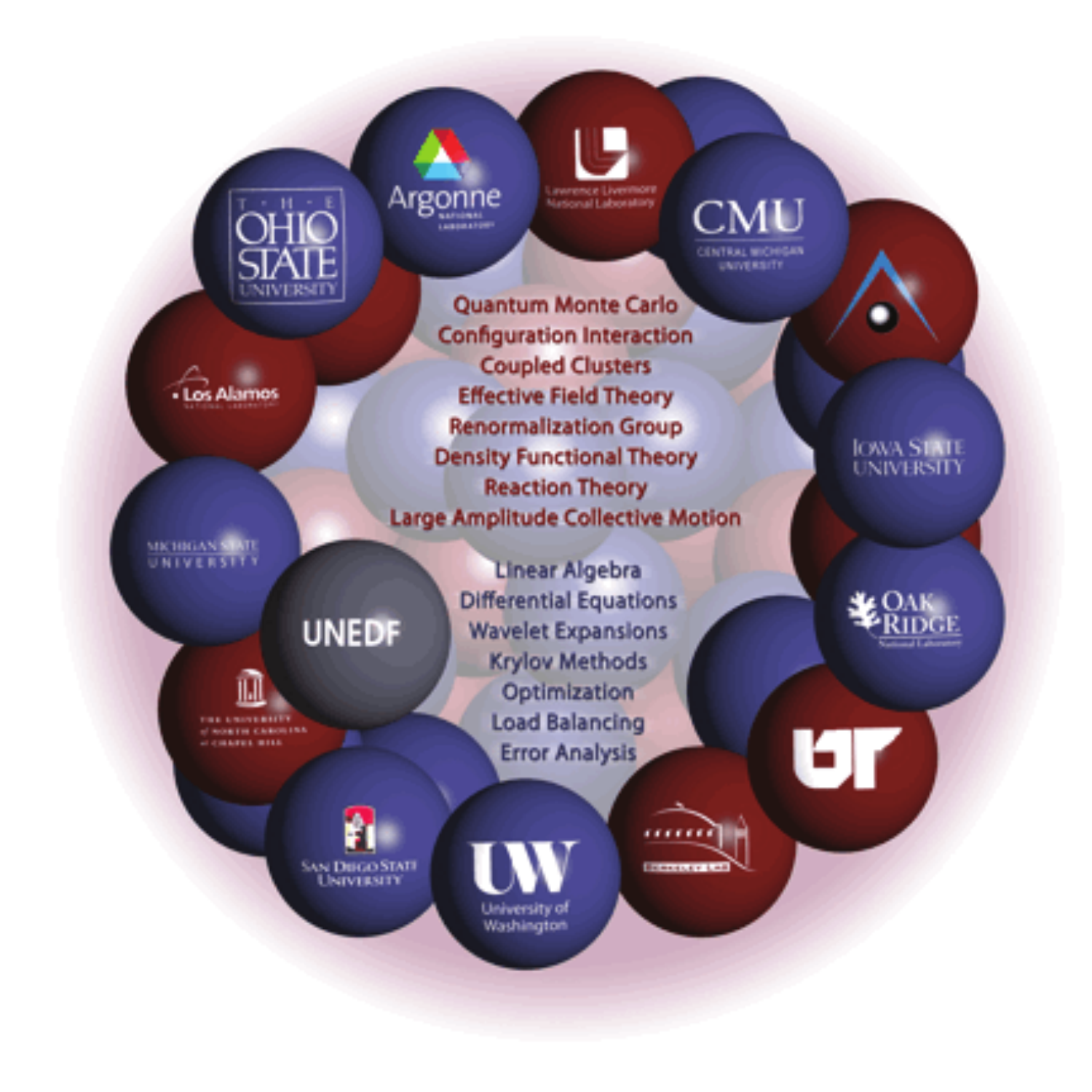}
\end{center}
\vspace{-20pt}
\caption{\small{UNEDF involves over 50 researchers from 9 universities and 7 national laboratories. Annually, it provides training to about 30 young researchers (postdocs and students).}}
\vspace{-20pt}
\label{fig1}
\end{wrapfigure}
Nuclear physicists are working toward a unified description of nuclei to transform current models into predictive capability; in particular, allowing reliable extrapolations into regions and regimes that are not accessible by experiments. Ultimately, this would allow for accurate predictions of nuclear reactions, with significant impact towards the development of advanced fission reactors and fusion energy sources, and in industrial and medical innovations via use of stable isotopes and radioisotopes.

The UNEDF collaboration of nuclear theorists, applied mathematicians, and computer scientists is making significant strides toward realizing this goal through a comprehensive study of all nuclei built on the latest advances in nuclear theory and scientific computing. UNEDF, which stands for ``Universal Nuclear Energy Density Functional'', is a five-year SciDAC project \cite{UNEDF}.  The SciDAC program has provided the opportunity for applied mathematicians and computer scientists to work collaboratively with physicists to develop advanced algorithms and tools and effectively utilize high-performance computing resources to achieve scientific breakthroughs in the low-energy nuclear many-body problem. An added benefit of the UNEDF project was the realization of new physics collaborations, identified through shared computational methods and needs.

Under UNEDF, collaborators develop and interconnect the most accurate knowledge of the strong nuclear interaction, high-precision theoretical approaches, scalable algorithms, and high-performance computing tools and libraries to enable scientific discoveries using leadership-class computing resources.  Working towards a predictive theory, the UNEDF project emphasizes the verification of methods and codes, the estimation of uncertainties, and assessment of results. Here we present some significant milestones achieved through the UNEDF collaborative effort and the future outlook; more details and references can be found at the unedf.org website.

\section{HPC Enhances \textit{Ab Initio} Nuclear Structure Calculations}
\label{sec1}
\textit{Ab initio}, or first principles, nuclear structure calculations have made major advances under UNEDF toward effectively utilizing high-performance computing resources and transforming to meet the challenges posed by emerging architectures.  \textit{Ab initio} techniques provide a fine grain method for studying nuclei and the nuclear interaction, but often come with a high computational cost.  They are necessary to the UNEDF effort by providing ``control-data" to constrain more general functionals and test candidate energy density functionals (EDF) .  UNEDF collaborators continue to scale \textit{ab initio} nuclear structure simulations and perform the largest and most accurate calculations currently possible on both Jaguar (ORNL) and Intrepid (ANL).

\begin{figure}[h]
\begin{minipage}{15pc}
\includegraphics[width=15pc]{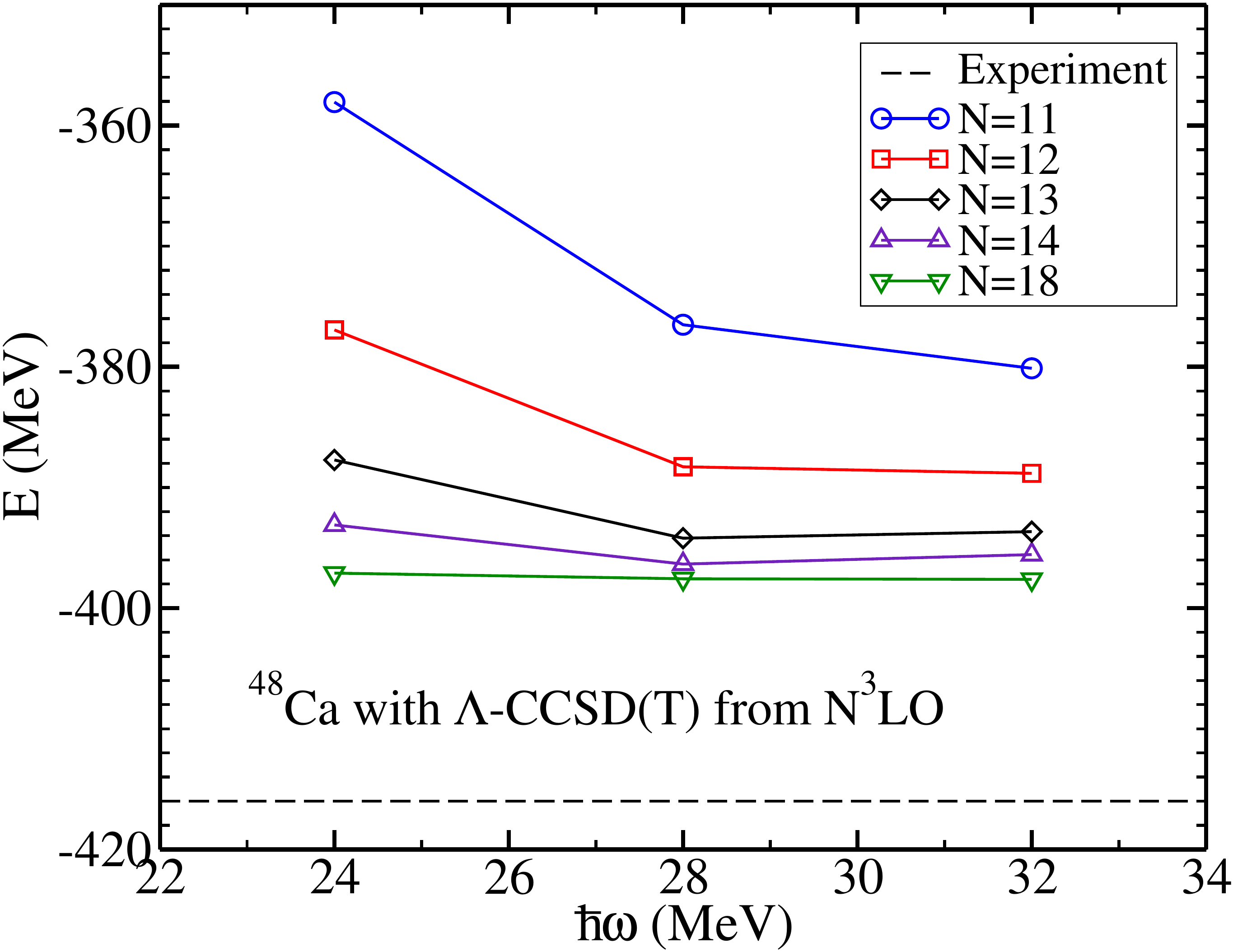}
\vspace{-10pt}
\caption{\small{NUCCOR calculations for medium-mass nuclei - the binding energy of $^{48}$Ca  as a function of  $\hbar \omega$ and the size of the model space \cite{HAGEN}.}}\label{fig2}
\end{minipage}
\hspace{3pc}
\begin{minipage}{17pc}
\vspace{-10pt}
\includegraphics[width=17pc]{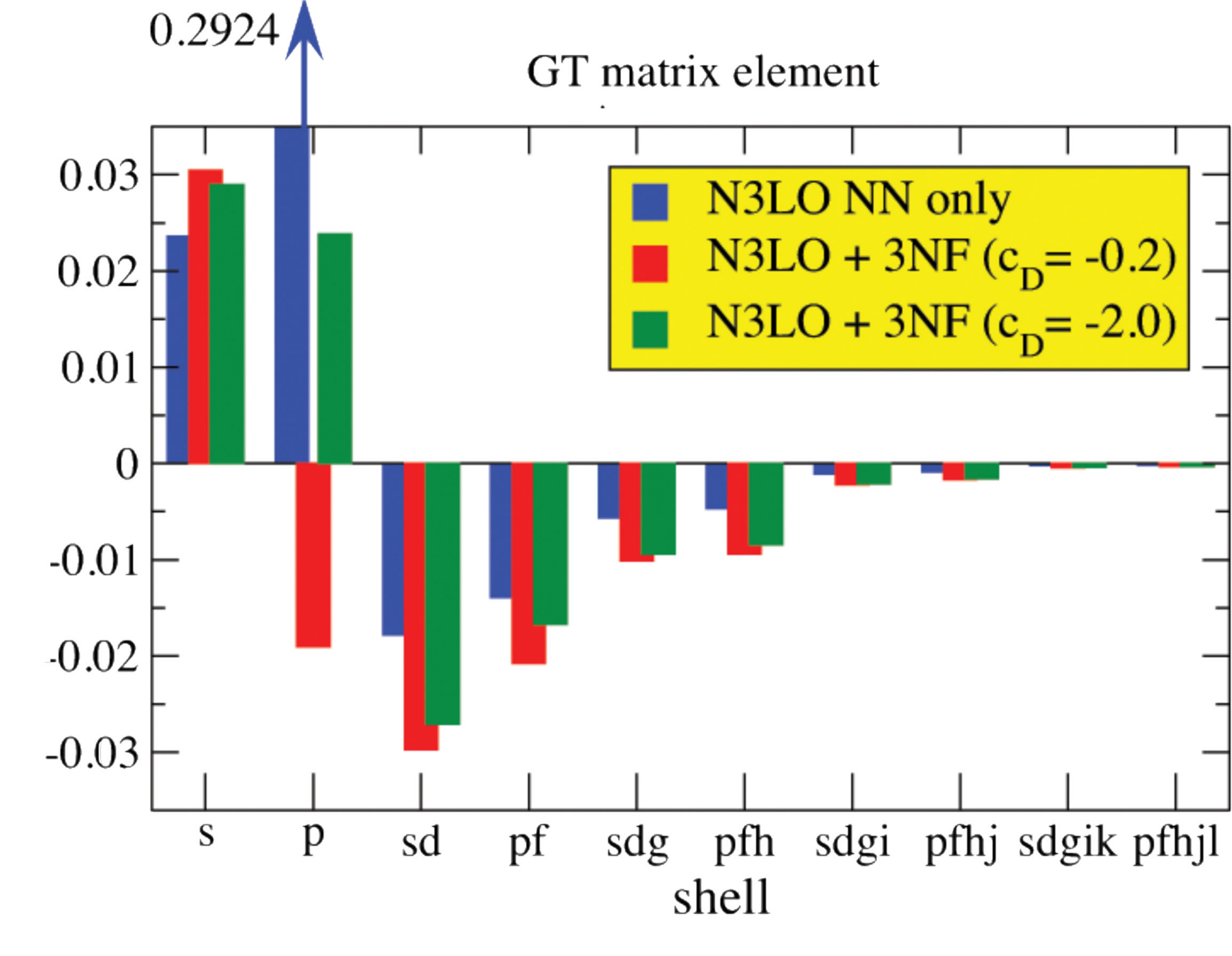}
\vspace{-20pt}
\caption{\small{MFDn simulation shows three-body forces are necessary to explain the anomalously long half-life of isotope $^{14}$C used in carbon dating \cite{MFDn}.}}\label{fig3}
\end{minipage}
\end{figure}

\vspace{-5pt}
Examples of UNEDF-directed advances include development of the Asynchronous Dynamic Load Balancing (ADLB) software library by using Green's Function Monte Carlo (GFMC) calculations as a test bed. The ADLB library has enabled GFMC to run efficiently on over 100,000 cores on Intrepid \cite{LUSK}.  Utilizing Jaguar, UNEDF applications \textit{Nuclear Coupled-Cluster - Oak Ridge} (NUCCOR) and \textit{Many Fermion Dynamics-nuclear} (MFDn) have undergone considerable code and algorithm development.  Improvements include implementation of a hybrid MPI and OpenMP approach for efficient memory management, memory aware algorithms, and integration of libraries and tools to enable further scaling for higher precision calculations.

Recent scientific breakthroughs include using NUCCOR to calculate medium-mass nuclei such as $^{48}$Ca shown in Fig.~\ref{fig2}  \cite{HAGEN}. Results show chiral nucleon-nucleon interactions perform remarkably well and the 400keV per nucleon missing binding energy in $^{48}$Ca can most likely be attributed to chiral three-nucleon (NNN) forces \cite{HAGEN}.  Another recent highlight explains the useful but anomalously long lifetime of $^{14}$C by identifying the critical role of the NNN force in its beta decay seen in Fig.~\ref{fig3} \cite{MFDn}.  These calculations involve diagonalization of a Hamiltonian matrix of dimension two billion using 214,668 cores on the Jaguar supercomputer at ORNL under a Petascale Early Science project using roughly 30 million core-hours.

\section{Advanced Algorithms and Tools Define New Generation EDF}
\label{sec3}
\begin{figure}
\begin{center}
\includegraphics[width=0.8\textwidth]{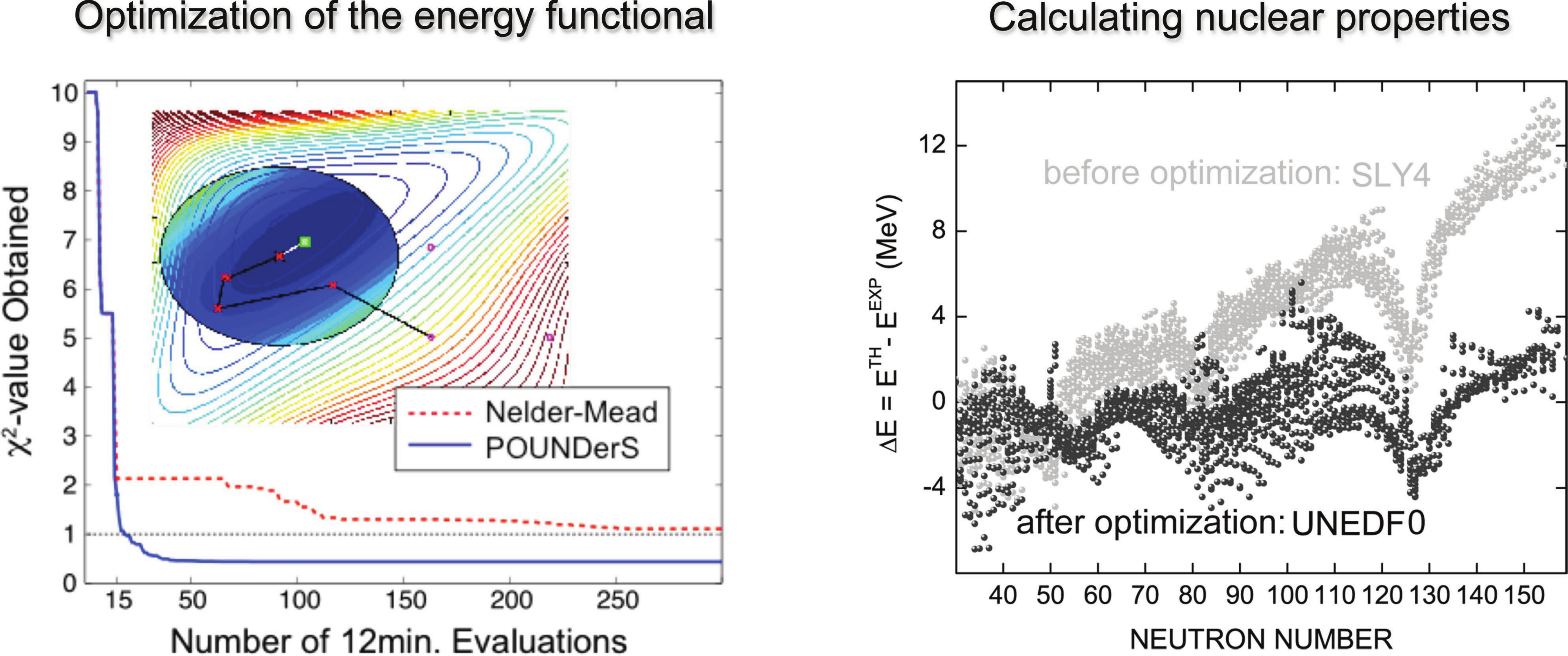}
\end{center}
\vspace{-10pt}
\caption{\small{The optimization algorithm POUNDerS yields dramatic computational savings over alternative optimization methods (left).  The resulting parameterization UNEDF0 obtained by using the POUNDerS algorithm provides a baseline of nuclear ground-state properties to compare with future functionals (right) \cite{UNEDF0}.}}
\label{fig4}
\vspace{-10pt}
\end{figure}

The UNEDF project has devoted considerable effort to develop and improve the algorithmic and computational infrastructure needed to optimize candidate EDFs.  These developments have resulted in new optimization tools which are broadly available to other science domains.  An example is the derivative-free  optimization algorithm POUNDerS, developed by a UNEDF team of applied mathematicians,  which provides a computational savings over other methods, as shown in Fig.~\ref{fig4}, and greatly improves the time to solution to test candidate EDFs. Using POUNDerS, the resulting parameterization of existing data yielded UNEDF0 which sets a solid baseline of nuclear ground-state properties to compare with future functionals \cite{UNEDF0}. Continuing work involves utilizing this approach to study new hybrid functionals with microscopic input from chiral effective field theory \cite{DME}.

\begin{figure}
\begin{center}
\includegraphics[width=0.8\textwidth]{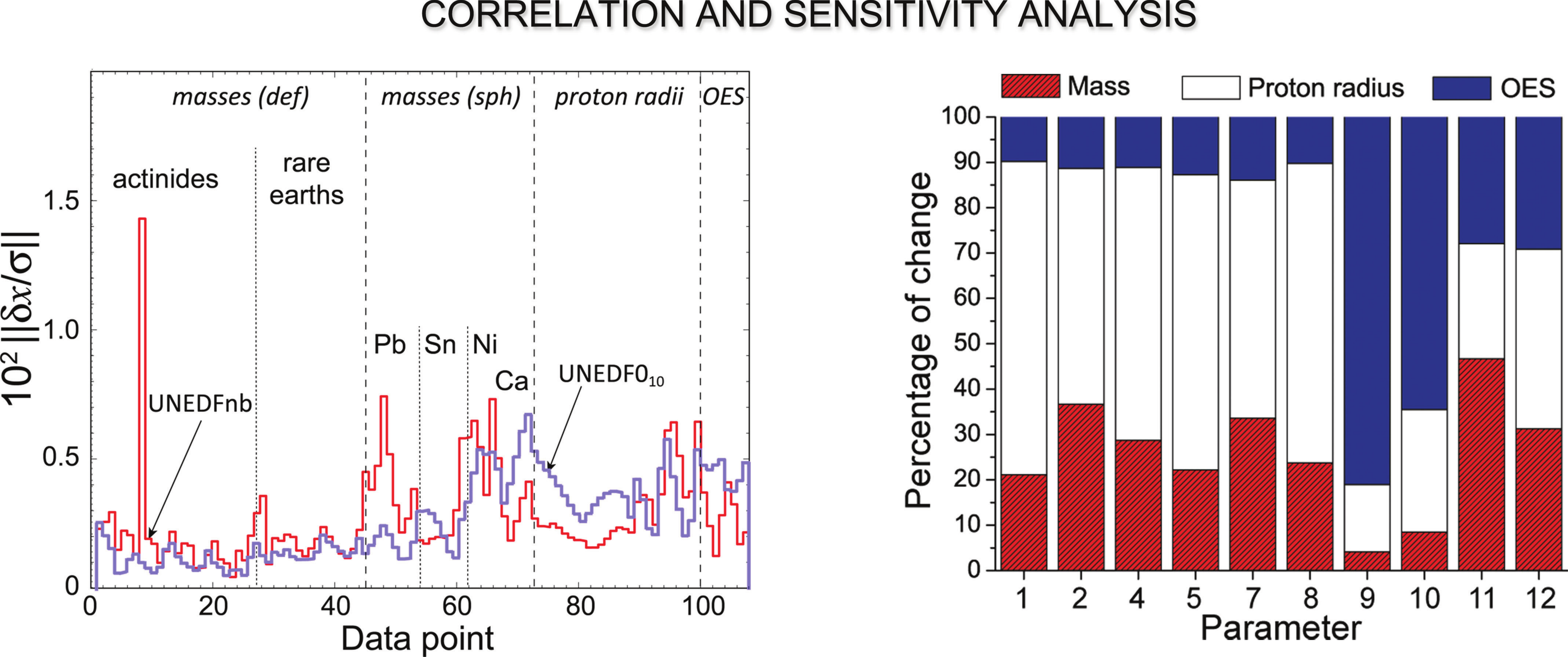}
\end{center}
\vspace{-10pt}
\caption{\small{Statistical tools are used to deliver uncertainty quantification and error analysis for theoretical studies as well as for the assessment of new experimental data \cite{UNEDF0}.}}
\label{fig5}
\vspace{-15pt}
\end{figure}

These new tools allow  a consistent method for uncertainty quantification and correlation analysis to estimate errors and significance as a first step towards a formal process for future verification and validation.   Inclusive in the UNEDF project is the development and application of statistical tools, particularly important for directing future experiments by providing analysis of the significance of new experimental data. For example, the sensitivity of two optimized functionals to particular data is shown in Fig.~\ref{fig5} \cite{UNEDF0}. Such capabilities have not been previously available in the low-energy nuclear theory community, but are increasingly more important as new theories and computational tools are applied to new nuclear systems and to conditions that are not accessible to experiment.

\section{Massively Parallel Algorithms Use Cold Atoms as a Testing Ground}
\label{sec4}
\begin{figure}[h]

\begin{minipage}{16pc}
\includegraphics[width=16pc]{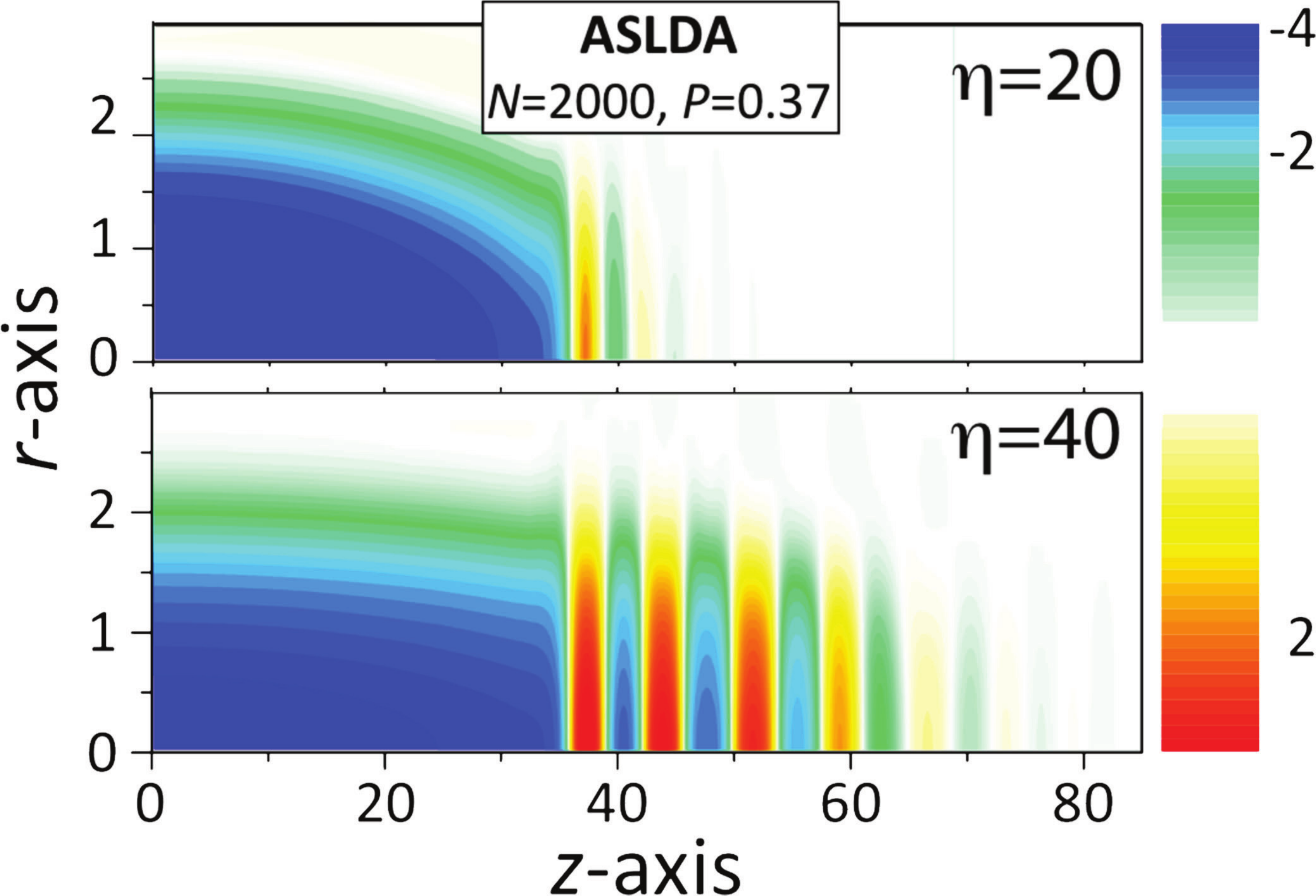}
\caption{\small{ASLDA simulation with strongly interacting spin-imbalanced atomic gases in extremely elongated traps \cite{PEI}.}}
\label{fig6}
\end{minipage}\hspace{4pc}
\begin{minipage}{17pc} 
~\\~\\~\\
\includegraphics[width=17pc]{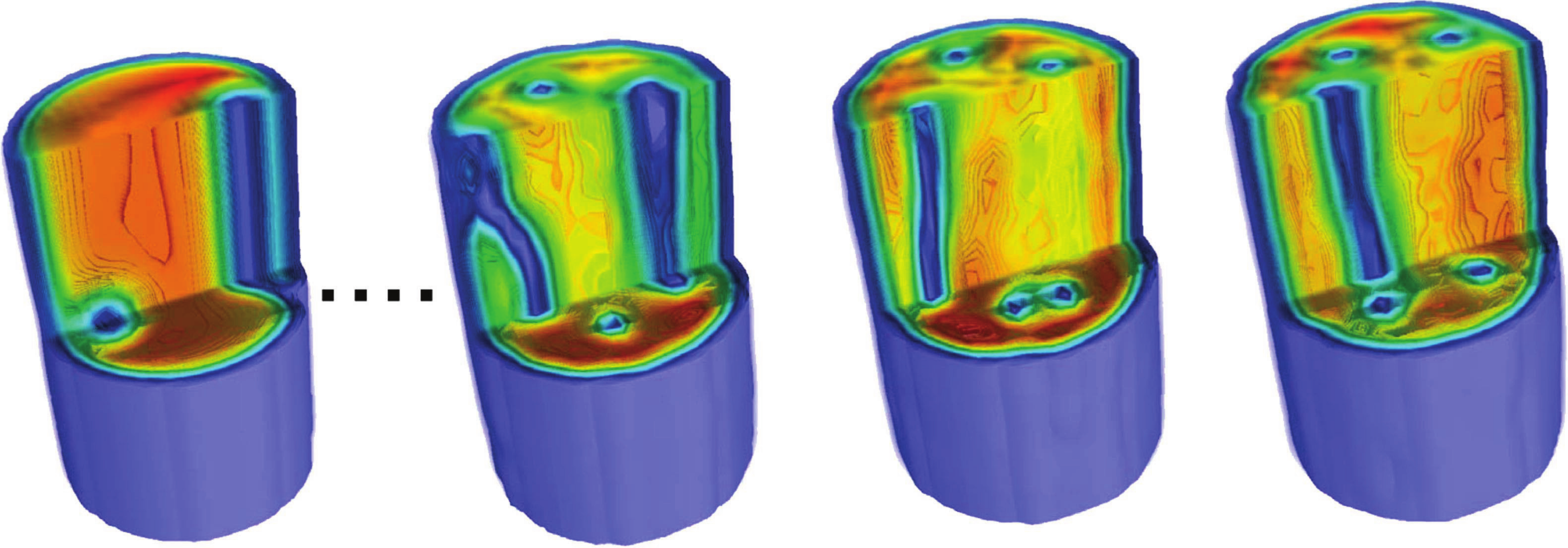}
\vspace{1.6pc}
\caption{\small{ TDSLDA simulation shows the ball and rod excitation of a unitary Fermi gas with vortex formation \cite{BULGAC}.}}
\label{fig7}
\end{minipage}
\end{figure}

UNEDF theorists have made important contributions to the study of strongly coupled superfluid systems such as ultracold Fermi atoms, which show many similarities to the cold nuclear matter found in the crust of neutron stars. Cold atoms make excellent laboratories for testing and improving the computational methods to be used for nuclei.  Cold atom systems also allow for predictions of superfluid density functional theory that are testable against experiment.

UNEDF developments using cold atoms as a testing ground include adding new algorithms to existing applications such as adapting the antisymmetric superfluid local density approximation (ASLDA) to an existing massively parallel nuclear DFT code with strongly interacting spin-imbalanced atomic gases in extremely elongated traps seen in Fig.~ \ref{fig6} \cite{PEI}.  Another major UNEDF development is implementation of the time-dependent superfluid local density approximation (TDSLDA) on a 3D spatial lattice \cite{BULGAC}. Unlike previous methods, the UNEDF implementation eliminates the need for matrix operations,  allowing it to accommodate a basis set that is 2-3 orders of magnitude larger than alternate approaches.  Calculations  \cite{BULGAC} were performed using 97\% of Jaguar to simulate the unitary gas including the vortex formation shown  in Fig.~\ref{fig7}. While still exploratory, these first-time simulations of this kind for fermion superfluids serve as proof of principle for an eventual treatment of fission.

\section{HPC Empowers A New Era for Nuclear Reaction Theory}
\label{sec5}

\begin{wrapfigure}{l}{0.4\textwidth}
\begin{center}
\includegraphics[width=0.4\textwidth]{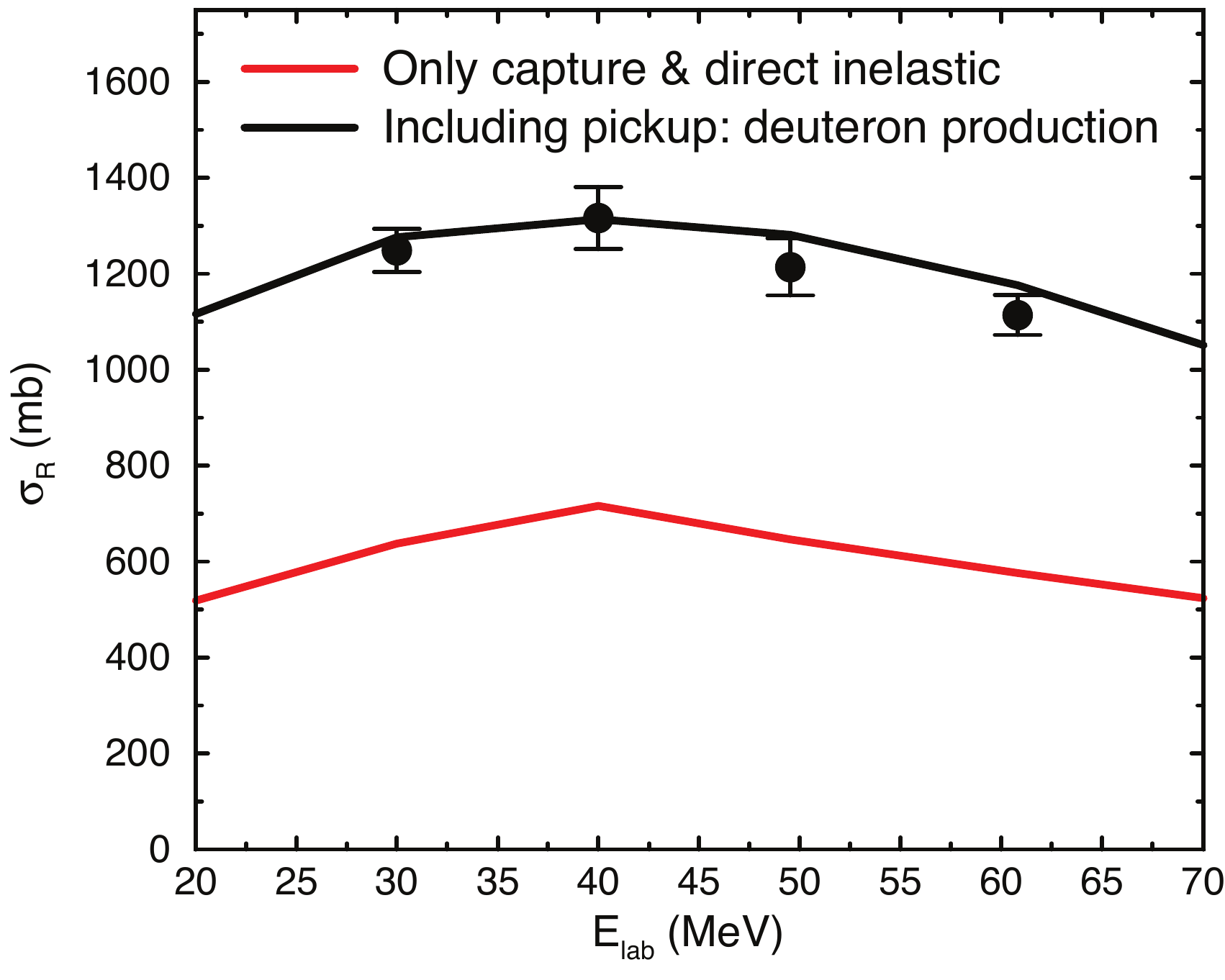}
\end{center}
\vspace{-20pt}
\caption{\small{New methods for calculating reaction cross-sections (black) show good agreement with experimental data \cite{IAN}.}}
\label{fig8}
\end{wrapfigure}

One of the principal aims of the UNEDF project is to calculate nucleus-nucleon reactions crucial for astrophysics, nuclear energy, radiobiology, and national security, for which extensions of standard phenomenology is insufficient.  Under the UNEDF effort, neutron reactions on heavier nuclei are being modeled using DFT results to predict not just bound states, but also scattering states for nucleons. As shown in Fig.~\ref{fig8}, the calculated reaction cross sections \cite{IAN} agree very well with experimental data.  

For the first time, a complete microscopic calculation using realistic EDFs can be used to predict reaction observables with low incident energy \cite{IAN}.   This technology provides the basis for future calculations of unstable species outside the range of experiment. 

Another important capability for reactions is the calculation of level densities which provides insight to the interactions inside the system. A new proton-neutron algorithm for the parallel JMoments code was recently designed and implemented, which scales to tens of thousands of cores, and greatly increases its overall performance. This development opens the door to calculating accurate nuclear level densities and reaction rates for a large class of nuclei \cite{IAN2}.

\vspace{70pt}
\section{Future Outlook}
\label{sec6}
The UNEDF collaboration is fertile ground for new and continuing growth between applied mathematics, computer science and nuclear physics.  Over the last five years, the collaboration has established cross-disciplinary working relationships to facilitate future efforts and has matured to adequately address new challenges in verification and validation, workflow, visualization, and new programming models with changing architectures.  The UNEDF project will continue to develop key computational codes and algorithms for reaching the goal of solving the nuclear quantum many-body problem, thus paving the road to the comprehensive description of the atomic nucleus and nucleonic matter.

\section{Acknowledgments}

The UNEDF SciDAC Collaboration is supported by the U.S. Department of
Energy (DOE) under grant No. DOE-FC02-09ER41583. This work was also supported
by DOE Contract Nos.~DE-FG02-96ER40963 (University of Tennessee) and DE-AC02-06CH11357 (Argonne).

\end{document}